\documentclass[journal, twocolumn]{IEEEtran}
\usepackage{amsmath,amsfonts,amsxtra,amssymb,latexsym,amscd,amsthm,mathrsfs,bm}
\DeclareUnicodeCharacter{2212}{-}
\usepackage{graphicx}
\usepackage{hyperref}
\usepackage{xcolor}
\usepackage{cite}
\usepackage{cuted}
\usepackage[small]{caption}
\usepackage{algorithm}
\usepackage{multirow}
\usepackage{comment} 
\usepackage{lipsum}
\usepackage{mathtools}
\usepackage{multicol}

\usepackage{algpseudocode}

\title{\LARGE Coordinate Interleaved Faster-than-Nyquist Signaling}
\author{
{{Adem Cicek, Enver Cavus, Ebrahim Bedeer, Ian Marsland, and Halim~Yanikomeroglu}}
\thanks{A. Cicek and E. Cavus are with Ankara Yildirim Beyazit University, Ankara, Turkey.
Emails: ademcicek@karatekin.edu.tr, ecavus@ybu.edu.tr. E. Bedeer is with University of Saskatchewan, SK, Canada. Email: e.bedeer@usask.ca, and H. Yanikomeroglu and I. Marsland are with Carleton University, ON, Canada. Emails: \{halim, ianm\}@sce.carleton.ca. \\ 
}}

\begin{document}
\maketitle



\begin{abstract}
Faster-than-Nyquist (FTN) signaling is an attractive transmission technique which accelerates data symbols beyond the Nyquist rate to improve the spectral efficiency; however, at the expense of higher computational complexity to remove the introduced intersymbol interference (ISI). In this work, we introduce a novel FTN signaling transmission technique, named coordinate interleaved FTN (CI-FTN) signaling that exploits the ISI at the transmitter to generate constructive interference for every pair of the {counter-clockwise} rotated binary phase shift keying (BPSK) data symbols. In particular, the proposed CI-FTN signaling interleaves the in-phase (I) and the quadrature (Q) components of the {counter-clockwise} rotated BPSK symbols to guarantee that every pair of consecutive symbols has the same sign, and hence, has constructive ISI. At the receiver, we propose a low-complexity detector that makes use of the constructive ISI introduced at the transmitter. Simulation results show the merits of the CI-FTN signaling and the proposed low-complexity detector compared to conventional Nyquist and FTN signaling.
\end{abstract}

\begin{IEEEkeywords}
Coordinate interleaving, constructive inter-symbol interference, low-complexity detection, faster-than-Nyquist.
\end{IEEEkeywords}

\IEEEpeerreviewmaketitle

\section{Introduction}\label{sec:intro}
\IEEEPARstart{L}{imited} spectrum resources require the use of novel spectral-efficient transmission techniques to support the increasing demand for high data rates. Faster-than-Nyquist (FTN) signaling is a promising spectral-efficient transmission technique that accelerates the transmit data symbols beyond the Nyquist limit and hence introduces a controlled intersymbol interference (ISI). 

In the literature, different low-complexity methods are adopted on the transmitter or receiver sides to handle the ISI of FTN signaling. 
At the receiver, the received symbols corrupted by the ISI are recovered with detection methods such as truncated Viterbi and 
M-BCJR algorithms \cite{Prlja6241379}, frequency-domain equalization (FDE)-assisted FTN receiver architecture \cite{response_MMSE_FDE}, low-complexity symbol-by-symbol based detector \cite{Bedeer7886296}, reduced-complexity M-BCJR algorithm based on the Ungerboeck observation model \cite{UNGERBOECK8114271}, and sum-product detector with deep learning approach\cite{Liu9458230}. At the transmitter side, Tomlinson-Harashima precoded (THP) FTN signaling system with a detector using an iterative successive interference cancellation (SIC) scheme is proposed in \cite{Wang8947975}. 
An FTN signaling with linear precoding that combines cyclic prefix, suffix, and discrete Fourier transform with existing linear precoding algorithms is introduced in \cite{Li8954896}. The precoding method has lower complexity than the non-linear counterparts such as THP; however, adding a prefix and suffix causes a reduction in the spectral efficiency. Also, Cholesky-decomposition aided precoding scheme with lower complexity and decoding matrix storage space consumption is applied for FTN signaling in  \cite{Yuan9352013}.

The techniques in the literature consider ISI as an undesirable effect and try to minimize the effect of ISI to achieve better performance. In an FTN system, the effect of the ISI on a given data symbol can be either constructive or destructive. Fig. 1 examplifies constructive and destructive interference effects for a simple case of the transmission of two consecutive binary phase shift keying (BPSK) symbols of $(+1, +1)$ and $(+1, -1)$, respectively. In other words, it shows that if we sample the first signal at $t=0$ (red line) or the second signal at $t=\tau T$ (the blue one) at the receiver, the neighboring signal may change the amplitude of the sampled symbol that will decrease or increase as in (a) and (b), respectively, depending on the sign. Therefore, at the transmitter, if more constructive interference can be generated by ordering the same signal symbols consecutively, the generated constructive effect might help to boost the performance of the FTN system.

\begin{figure}[t]
\centering
\includegraphics[scale=0.45]{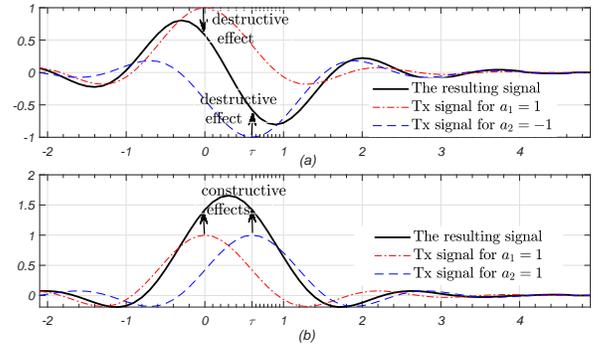}
\caption{Illustration of the destructive \textit{(a)} and constructive \textit{(b)} ISI effects for $\tau = 0.6$} 
\label{cons_dest_presentation}
\vspace{-4.6mm}
\end{figure}

In this work, we introduce a simple FTN transmission method, which benefits from the ISI present on FTN signaling to improve the system performance with a negligible complexity increase. Inspired by the idea of coordinate interleaving proposed in \cite{Boutros681321}, we first rotate the BPSK symbols counter-clockwise and then interleave the real and imaginary components of the rotated symbols such that the consecutive symbols possess the same signs. Having signal pairs with the same polarity guarantees constructive interference between the adjacent FTN symbols, which contributes to the symbol power by reducing destructive ISI from the adjacent symbol, and hence improves the performance. Moreover, proposed coordinate interleaved FTN (CI-FTN) signaling also reduces the total destructive ISI within a single block of symbols and improves the overall detection performance. Please note that the BPSK CI-FTN signaling can be viewed as a combination of the QPSK FTN signaling and repetition coding, where the symbols of the same sign are back to back. In this way, consecutive symbols with the same sign allow ISI reduction and a very low complexity detector design in the FTN signaling environment. Simulation results show the performance merits of the CI-FTN signaling with the low-complexity detector in terms of lower bit error rates when compared to the optimal detection of conventional Nyquist and FTN signaling.

The remainder of this paper is organized as follows. Section \ref{sec:model} presents the proposed CI-FTN signaling system model, while an ISI analysis of the proposed technique is discussed in Section \ref{sec:analysis}. A simple and low-complexity detector for the CI-FTN technique is introduced in Section \ref{sec:detector}. Section \ref{sec:simulation} presents the simulation results, and the paper is concluded in Section \ref{sec:conc}.
\section{CI-FTN Signaling System Model}{\label{sec:model}}

\begin{figure}[t]
\centering
\includegraphics[width=0.5\textwidth]{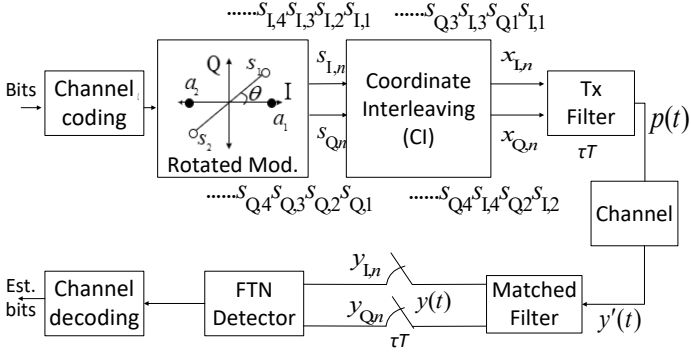}
\caption{CI-FTN signaling system model.} \label{system_model}
\end{figure}

The system model of the proposed CI-FTN signaling is shown in 
Fig. \ref{system_model}. At the transmitter side, $K$-bit streams are encoded to $N$-length codewords in channel encoder with the coderate $K/N$ and then the encoded bits are mapped to BPSK data symbols $a_{n}, n = 1, ..., N,$ each of energy $E_b$, that are \textit{counter-clockwise} rotated by an angle $\theta = \pi/4$ as can be seen from rotated modulation block in Fig. \ref{system_model}.  
The {counter-clockwise} rotated BPSK symbols $s_{n}, n = 1, ..., N,$ are given as 
\begin{equation}\label{rotation}
\begin{aligned}
s_{n} &= a_{n}\exp\left(j\theta\right) = s_{\text{I},n} + j s_{\text{Q},n}, \quad n = 1, ..., N,
\end{aligned}
\end{equation}
where $s_{\text{I},n}$ and $s_{\text{Q},n}$ represent the I and Q components of the rotated $n$th BPSK data symbol, respectively, and $j = \sqrt{-1}$. After rotation, we perform CI on every two consecutive symbols $s_n$ and $s_{n+1}$, $n = 1, ..., N$, to produce two new symbols $x_n$ and $x_{n+1}$ that consist of the in-phase and quadrature components of $s_n$ and $s_{n+1}$, respectively. CI can be formally expressed as
\begin{equation}\label{interleaving}
\begin{aligned}
x_{n}  = f(s_n) = \Bigg\{\begin{array}{lr}
        s_{\text{I},n} + js_{\text{I},n+1}, & \text{if } n  \text{ is odd}\\
    s_{\text{Q},n-1} + js_{\text{Q},n}, &  \text{if } n \text{ is even}\\
        \end{array},\\
\end{aligned}
\end{equation}
where $x_{n}$ is the $n$th interleaved symbol, and $f(\mathord{\cdot})$ is a function representing the interleaving operation. The CI operation described earlier ensures that we transmit the I and Q components of the odd numbered {counter-clockwise} rotated BPSK symbol using the in-phase channel, while sending the I and the Q components of the even numbered {counter-clockwise} rotated BPSK symbols on the quadrature channel. Hence, it is important that the rotation is done {counter-clockwise} to guarantee that the I and Q components of any BPSK data symbol will have the same sign. In other words, if the inputs to the CI on the in-phase and quadrature channels are $\{s_{\text{I},1}, s_{\text{I},2}, ..., s_{\text{I},N}\}$ and $\{s_{\text{Q},1}, s_{\text{Q},2}, ..., s_{\text{Q},N}\}$, respectively, then the outputs of the CI on the in-phase and quadrature channels are $\{s_{\text{I},1}, s_{\text{Q},1}, ..., s_{\text{I},N-1}, s_{\text{Q},N-1}\}$ and $\{s_{\text{I},2}, s_{\text{Q},2}, ..., s_{\text{I},N}, s_{\text{Q},N}\}$, respectively. Please note that the CI operation does not increase the number of time slots required to transmit the $N$ BPSK symbols, and it uses two dimensions for the BPSK modulation; hence, the spectral efficiency (SE) of BPSK CI-FTN is compared with QPSK Nyquist signaling.

Following the CI block, the interleaved symbols, $x_{n}$, $n = 1,..,N$ is transmitted through a unit energy root raised cosine (rRC) transmit filter every $\tau T$, where $\tau$ is the acceleration parameter and $T$ is the symbol duration. The transmit signal is expressed as
\begin{equation}\label{pulseshaping}
\begin{aligned}
p(t) = \zeta  \sum_{n=1}^{N} x_{n}h(t-n\tau T),
\end{aligned}
\end{equation}
where  $\zeta$ is a normalization factor to ensure equal transmit energy of the CI-FTN signaling when compared to conventional FTN signaling. 
In this work, we consider an additive white Gaussian noise (AWGN) channel; hence, the received signal $y'(t)$ can be given as
\begin{equation}\label{passingtruchannel}
\begin{aligned}
y'(t) = p(t) + w(t),
\end{aligned}
\end{equation}
where $w(t)$ is a zero mean 
complex valued Gaussian noise with a covariance of $\sigma^{2} \boldsymbol{I}$, where $\boldsymbol{I}$ is identity matrix.
At the receiver side, the received signal is passed through a filter matched to the rRC transmit filter. The matched filter output is written as
\begin{equation}\label{matched_filt_output}
\begin{aligned}
y(t) = \zeta \sum_{n=1}^{N} x_{n} g(t-n\tau T) + q(t),
\end{aligned}
\end{equation}
where $ g(t) = \int_{}^{} h(x')h(x'-t) dx'$, $ q(t) = \int w(x')h(x'-t)dx'$. Since we transmit the symbols at a rate of $1/(\tau T)$, the output of the matched filter is sampled at the same rate. Next, the sampled signal is processed using the FTN detector block, and its soft output is input into the channel decoder. Finally, the decoder decodes them and then outputs the bit streams.

Now we formulate the optimal detection rule for the proposed CI-FTN system. The sampled signal in \eqref{matched_filt_output} can be expressed in a vector form as 
\begin{equation}\label{sampler_out_vector_form}
\begin{aligned}
\boldsymbol{y} = \zeta \boldsymbol{Gx} + \boldsymbol{q},
\end{aligned}
\end{equation}
where $\boldsymbol{G}$ is the $N \times N$ ISI matrix, and $\boldsymbol{x} = f(\boldsymbol{s}) = f(\boldsymbol{a} \exp(j\theta))$ is the vector of the interleaved symbols, and $\boldsymbol{q}$ is an $N \times 1$ zero-mean sampled noise vector with covariance of $\sigma^{2} \boldsymbol{G}$. Assuming that $\boldsymbol{G}$ is invertible, \eqref{sampler_out_vector_form} can be also written as 
\begin{equation}\label{re-sampler_out_vector_form}
\begin{aligned}
\boldsymbol{z} = \frac{1}{\zeta} \boldsymbol{G}^{-1}\boldsymbol{y} = \boldsymbol{x} + \frac{1}{\zeta} \boldsymbol{G}^{-1}\boldsymbol{q},\\
\end{aligned}
\end{equation}
where the left hand side $\frac{1}{\zeta} \boldsymbol{G}^{-1}\boldsymbol{y}$  can be thought of as a Gaussian random variable $\boldsymbol{z}$ with covariance $\Delta =  $
$\sigma^{2} \frac{1}{\zeta^2} \boldsymbol{G^{-1}}$ and mean $\boldsymbol{x}$. Hence, probability density function (pdf) of $\boldsymbol{z}$ is written as 
\begin{equation}\label{pdf}
\begin{aligned}
P(\boldsymbol{z}|\boldsymbol{x}) = \frac{1}{\sqrt{(2\pi)^{N}|\Delta|}}\text{exp}(-\frac{1}{2}(\boldsymbol{z}-\boldsymbol{x})^\text{T}\Delta^{-1}(\boldsymbol{z}-\boldsymbol{x})),
\end{aligned}
\end{equation}
where $|\Delta|$ is the determinant of $\Delta$. Finding a sequence $\boldsymbol{\hat{x}}$ to maximize \eqref{pdf} is known as the maximum likelihood sequence estimation (MLSE). Since maximizing \eqref{pdf} is equivalent to minimizing the exponent part in \eqref{pdf}, the MLSE problem for the CI-FTN can be expressed as

\begin{equation}\label{MLSE_short}
\begin{aligned}
\mathcal{OP}_\text{MLSE}: \boldsymbol{\hat{a}} &=
\text{arg}\underset{\boldsymbol{a}\in D}{\text{min}} \bigg(\Big(\boldsymbol{z}-f\big(\boldsymbol{a}\exp(j\theta)\big)\Big)^\text{T}\Delta^{-1}\\
&\Big(\boldsymbol{z}-f\big(\boldsymbol{a}\exp(j\theta)\big)\Big)\bigg).
\end{aligned}
\end{equation}

Finding the optimal solution to the MLSE in \eqref{MLSE_short} is of high-complexity. In Section IV, we discuss a low-complexity detection technique for CI-FTN signaling.

\section{ISI on the CI-FTN Technique} \label{sec:analysis}
This section examines the effect of coordinate interleaving on ISI that affects the symbol. It shows how CI-FTN reduces destructive ISI in a single symbol block at different $\tau$ values compared to traditional FTN signaling.

As explained in the previous section, after the CI operation at the transmitter the odd-numbered and even-numbered {counter-clockwise} rotated BPSK symbol components exist on the in-phase and quadrature channels, respectively. This ensures that every two interleaved consecutive symbol components will have the same sign and thus decreasing the power loss at the receiver side compared to conventional FTN signaling  as the CI-FTN reduces destructive interference, which may cause the detector to make a wrong  decision. In order to better understand the advantages of the proposed CI-FTN method, we will give a numerical example in the absence of channel noise. Let us consider transmitting the following sequence of $6$ BPSK symbols $[1 \; -1 \; 1 \; -1 \; 1 \; 1]$ and examine the ISI effects on the third symbol for the case of $\tau=0.6$. In conventional FTN signaling, the real part of the sequence after matched filter is $[0.5250 \; 0.1154 \; -0.3340 \; -0.0988 \; 1.0166 \; 1.4706]$. In this case, the third symbol gets a total ISI of $-1.3340$ and the resulting symbol value becomes $-0.3340$, which has a flipped sign compared to the transmitted symbol sign. On the other hand, when the CI-FTN method is used for the same sequence, we transmit [0.5794 - 0.5794i\;   0.5794 - 0.5794i\;   0.5794 - 0.5794i\; 0.5794 - 0.5794i\;   0.5794 + 0.5794i\; 0.5794 + 0.5794i]. Consequently, the I and Q components of the $3$ rd symbol, namely $x_{I,3}$ and $x_{I,4}$ according to Fig. \ref{sec:model} are now adjacent and constructive to each other, which helps to make a correct detection.
Thus, ordering the same sign components consecutively in CI-FTN technique reduces the individual ISI on each symbol component, which leads to a considerable reduction in the overall interference within a symbol block. Based on the worst-case sequence, Table \ref{worst_case_ISI} shows ISI values experienced by a symbol within a sequence of $2L$ symbols in conventional FTN and CI-FTN signaling for different values of $\tau$, where $L$ is the length of the single-sided ISI vector. For example, for $\tau=0.6$,
while a symbol can experience ISI up to $165\%$ of the magnitude of the symbol in conventional FTN signaling, a symbol component (I or Q) is exposed to ISI up to $47\%$ of the magnitude of the component in the CI-FTN technique at $\tau=0.6$. This significant ISI reduction in the CI-FTN case is achieved by exposing the consecutive components with the same sign. 

\begin{table}
    \centering
    \caption{ISI values at different values of $\tau$.}
    \label{worst_case_ISI}
    \begin{tabular}{|c|c|c|c|c|}
\hline
\multicolumn{1}{|c|}{}&\multicolumn{2}{|c|}{\multirow{2}{6em}{Conventional FTN signaling}}&\multicolumn{2}{|c|}{\multirow{2}{4em}{CI-FTN signaling}}\\
&\multicolumn{2}{|c|}{}&\multicolumn{2}{|c|}{}\\
\hline
\multirow{2}{1em}{$\tau$}&\multirow{2}{4em}{Symbol magnitude}
&\multirow{2}{4em}{ISI values}&\multirow{2}{4em}{ISI values}&\multirow{2}{5em}{Component magnitude}\\
\multicolumn{1}{|c|}{}&\multicolumn{1}{c}{}&\multicolumn{1}{|c|}{}
&\multicolumn{1}{c|}{}&\multicolumn{1}{c|}{}\\
\hline
0.9 & 1 & 0.5388 & 0.0464 & 0.5794 \\ 
\hline
0.8 & 1 & 0.9727 & 0.0649 & 0.5794 \\ 
\hline
0.7 & 1 & 1.2913 & 
0.1552 & 
0.5794 \\ 
\hline
0.6 & 1 & 1.6580 & 
0.2349 & 
0.5794 \\ 
\hline
0.5 & 1 & 1.8547 & 
0.3718 & 
0.5794 \\ 
\hline
0.4 & 1 & 2.8911 & 
0.7278 & 
0.5794 \\ 
\hline
\end{tabular}
\end{table}

\section{A Simple Detector for the CI-FTN Technique} \label{sec:detector}
In this section, we introduce a low-complexity detector for the proposed CI-FTN signaling, which strikes a balance between computational complexity and performance.

The proposed low-complexity detector exploits the fact that the I and Q components of each interleaved pair belong to the same BPSK data symbol, and hence, should have the same sign. 
As discussed earlier, the I and Q components of the rotated BPSK symbols $s_{k}$ and $s_{k+1}$ are interleaved with each other to produce two interleaved symbols $x_{k}$ and $x_{k+1}$ as described in \eqref{interleaving}. 
After the matched filter and sampler, the I components of the first two received samples $y_{\text{I},k}$ and $y_{\text{I},k+1}$ can be expressed as
\begin{equation}\label{received_signal1}
\begin{aligned}
y_{\text{I},k} &= \underbrace{\zeta G_{kk}\overbrace{x_{\text{I},k}}^{s_{\text{I},k}}}_{\text{I component of the $k^{\text{th}}$ symbol}} + \underbrace{ \zeta G_{k(k+1)}\overbrace{x_{\text{I},k+1}}^{s_{\text{Q},k}}}_{\text{constructive ISI}}\\
& + \underbrace{\zeta \sum_{i=1}^{L-1}x_{\text{I},k-i} G_{k(k-i)}}_{\text{ISI from previous components}} + \underbrace{\zeta \sum_{i=2}^{L}x_{\text{I},k+i} G_{k(k+i)}}_{\text{ISI from upcoming components}}\\
& + q_{\text{I},k.}
\end{aligned}
\end{equation}
\begin{equation}\label{received_signal2}
\begin{aligned}
y_{\text{I},k+1} &= \underbrace{\zeta G_{(k+1)(k+1)}\overbrace{x_{\text{I},k+1}}^{s_{\text{Q},k}}}_{\text{Q component of the $k^{\text{th}}$ symbol}} + \underbrace{\zeta G_{(k+1)k}\overbrace{x_{\text{I},k}}^{s_{\text{I},k}}}_{\text{constructive ISI}} \\
& + \underbrace{\zeta \sum_{i=2}^{L}x_{\text{I},k+1-i} G_{(k+1)(k+1-i)}}_{\text{ISI from previous components}}\\
& + \underbrace{\zeta \sum_{i=1}^{L-1}x_{\text{I},k+1+i} G_{(k+1)(k+1+i)}}_{\text{ISI from upcoming components}} + q_{\text{I},k+1,}
\end{aligned}
\end{equation}

Similarly, the received samples on the Q channel can also be written for the $y_{\text{Q},k}$ and $y_{\text{Q},k+1}$ signals. As can be seen from (\ref{received_signal1}) and (\ref{received_signal2}), the component interleaving guarantees that dominant ISI term that comes from the adjacent symbol within a pair of symbols is constructive. To estimate the transmitted symbol $s_k$, the proposed detector must jointly consider $y_{\text{I},k}$ and $y_{\text{I},k+1}$ as both samples correspond to the I and Q components of the same data symbol $s_k$. Similarly, the samples $y_{\text{Q},k}$ and $y_{\text{Q},k+1}$ are processed jointly to determine the transmitted symbol $s_{k+1}$.
\begin{strip}
\begin{equation}
\label{sum_received_signal1_2}
\begin{aligned}
y_{\text{I},k} + y_{\text{I},k+1}
&=\underbrace{\zeta G_{kk}\overbrace{x_{\text{I},k}}^{s_{\text{I},k}} + \zeta G_{(k+1)(k+1)}\overbrace{x_{\text{I},k+1}}^{s_{\text{Q},k}} + \zeta G_{k(k+1)}\overbrace{x_{\text{I},k+1}}^{s_{\text{Q},k}} + \zeta G_{(k+1)k}\overbrace{x_{\text{I},k}}^{s_{\text{I},k}}}_{\text{sum of the terms with the same sign as $k^{\text{th}}$ symbol}}
+\underbrace{\zeta \sum_{i=1}^{L-1}x_{\text{I},k-i} G_{k(k-i)} + \zeta \sum_{i=2}^{L}x_{\text{I},k+i} G_{k(k+i)}}_{\text{ISI}_{k} \text{sum of ISI from previous and upcoming comp. for $k^{\text{th}}$ sample}}\\ &+\underbrace{q_{\text{I},k}}_{\text{noise sample for $k^{\text{th}}$ sample}}
+\underbrace{\zeta \sum_{i=2}^{L}x_{\text{I},k+1-i} G_{(k+1)(k+1-i)} \zeta \sum_{i=1}^{L-1}x_{\text{I},k+1+i} G_{(k+1)(k+1+i)}}_{\text{ISI}_{k+1} \text{sum of ISI from previous and upcoming components for $(k+1)^{\text{th}}$ sample}}
+\underbrace{q_{\text{I},k+1}}_{\text{noise sample for $(k+1)^{\text{th}}$ sample}}.
\end{aligned}
\end{equation}
\end{strip}where $G_{\mu \nu}$ is an element of the ISI matrix $\mathbf{G}$ which represents the ISI effect of $\nu{\text{th}}$ symbol on the $\mu{\text{th}}$ symbol.
\begin{algorithm}[b]
\caption{Proposed CI-FTN detector}\label{alg:detector_for_ci-ftn}
\begin{algorithmic}
\State \textbf{Inputs:} $\textbf{\textit{y, G}}, L, N, \theta, \zeta$
\State \textbf{Outputs:} $\hat{\textbf{\textit{a}}}$
\State \textbf{Initialization:}
\State $k \leftarrow 1$
\State $n \leftarrow 1$
\For{$k < N$}
\State $\text{sumIQ}_{\text{k}} = y_{\text{I},k} + y_{\text{I},k+1}$
\For{$k-L \leq n \leq k+L$}
\State Calculate $\text{ISI}_{k}$, $\text{ISI}_{k+1}$
\State $n \leftarrow n+1$
\EndFor  
\State $\hat{a}_{k} \leftarrow \text{quantize}(\text{sumIQ}_{\text{k}} -\text{ISI}_{k+1} -\text{ISI}_{k})$
\State $k \leftarrow k+2$
\EndFor
\State \textbf{return} $\hat{\textbf{\textit{a}}}$
\end{algorithmic}
\end{algorithm}
As the components of a symbol are exposed to different levels of interference, there exists a diversity benefit at the receiver. To utilize this diversity, the simplest approach would be to add $y_{\text{I},k}$ and $y_{\text{I},k+1}$ before starting the detection process. In \eqref{sum_received_signal1_2}, it is shown that when added together, all the terms with the same sign as $s_k$ are combined to obtain a more reliable sample for the estimation of $s_k$.
The proposed detector for the CI-FTN technique is formally described in Algorithm \ref{alg:detector_for_ci-ftn}, where the detection process is started after receiving at least $L$ samples on each of the I and Q channels, where $L << N$ is the one sided ISI length. Please note that waiting for $L$ samples to start the detection process does not have any negative effect on latency, since system latency is usually determined by the block length of channel codes, which is generally much larger than $L$. 
In Algorithm 1, to detect the symbol $a_k$, first the two components of a symbol, $y_{\text{I},k}$ and $y_{\text{I},k+1}$, which are in the same pair and have the same sign, are summed together. Then, $L$ received samples on both sides of $s_{\text{I},k}$ and $s_{\text{Q},k}$, and the ISI on them is estimated. Note that the ISI consists of $\text{ISI}_{k+1}$ and $\text{ISI}_{k}$ which represent the total ISI amount from the previous and upcoming components outside of $x_{\text{I},k}$ and $x_{\text{I},k+1}$ component pair on ${k}$th and ${(k+1)}$th components, respectively. 
These ISI values are calculated using the quantized samples for the ISI from the upcoming samples, and the estimated symbols $\hat{a}_k$ for the ISI from the previous samples. Removing the ISI due to the previous and upcoming samples, which is seen from \eqref{received_signal1} and \eqref{received_signal2} clearly leaves only the constructive ISI effects at $n=k+1$ and $n=k$ on the symbol components.

The $k^{\text{th}}$ symbol on the I path is estimated as in \eqref{est_symbol_1}. Similarly, the $(k+1)^{\text{th}}$ symbol on the Q path can be estimated.
\begin{equation}\label{est_symbol_1}
\begin{aligned}
\hat{a}_{k} &= \text{quantize}\Big\{ 
y_{\text{I},k} + y_{\text{I},k+1} - \underbrace{\zeta \sqrt{\frac{1}{2}} \sum_{i=1}^{L-1} \hat{a}_{k-i} G_{k(k-i)}}_{\text{ISI from prev. sym. for $x_{\text{I},k}$ sample}}\\
&- \underbrace{\zeta \sqrt{\frac{1}{2}} \sum_{i=2}^{L} \text{quantize}\{y_{\text{I},k+i}\}  
G_{k(k+i)}}_{\text{ISI from upcoming symbols for $x_{\text{I},k}$ component}}\\
&- \underbrace{\zeta \sqrt{\frac{1}{2}} \sum_{i=1}^{L-1} \text{quantize}\{y_{\text{I},k+1+i}\} G_{(k+1)(k+1+i)}}_{\text{ISI from upcoming symbols for $x_{\text{I},k+1}$ component}}\\
&- \underbrace{\zeta \sqrt{\frac{1}{2}} \sum_{i=2}^{L} \hat{a}_{k+1-i} G_{(k+1)(k+1-i)}}_{\text{ISI from previous symbols for $x_{\text{I},k+1}$ component}}\Big\},
\end{aligned}
\end{equation}
where \text{quantize\{$\mathord{\gamma}$\}} rounds $\mathord{\gamma}$ to the nearest unrotated BPSK symbol.
\begin{table}[!b]
    \centering
    \caption{Hardware complexity.}
    \label{complexity}
    \begin{tabular}{|c|c|c|}
          \hline
 Algorithm & Complexity\\
 \hline
 The proposed detector & $\mathcal{O}(NL)$\\
\hline
  MMSE FDE \cite{response_MMSE_FDE} & $\mathcal{O}(NlogN)$\\
\hline
 BCJR \cite{Prlja6241379} & $\mathcal{O}(2^{L}N)$\\
\hline
\end{tabular}
\end{table}
When it comes to the complexity of the proposed detector, as seen from Table \ref{complexity}, the proposed detector has the complexity of $\mathcal{O}(NL)$ while the one of the BCJR detector \cite{Prlja6241379} is $\mathcal{O}(2^{L}N)$, meaning that its complexity increases exponentially as $L$ increases. The work in \cite{response_MMSE_FDE} has similar complexity to the proposed detector, but requires waiting for all the $N$ samples in the sequence for detection.
\section{Simulation Results}\label{sec:simulation}
\begin{figure}[!t]
\centering
\includegraphics[width=0.55\textwidth]{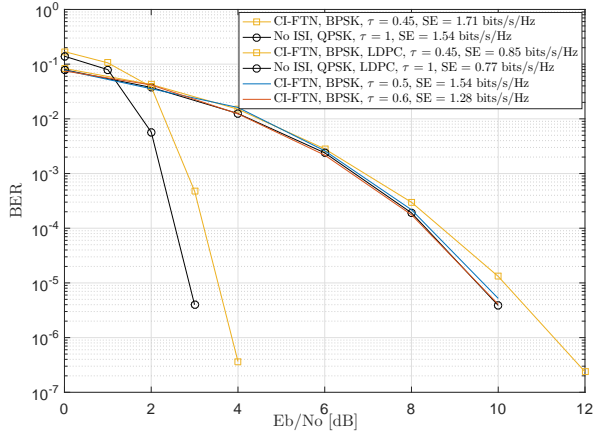}
\caption{Channel coded and uncoded performance of the proposed detector using the CI-FTN technique, $\alpha = 0.3$.}
\label{proposed_method_results}
\vspace{-2mm}
\end{figure}

\vspace{8pt}
In this section, we evaluate the performance of the proposed CI-FTN signaling using the proposed low-complexity detector and compare with different FTN detectors. 
The roll-off factor of the root raised cosine is set to  $\alpha = 0.3$. We also show the LDPC coded performance of the detector against Nyquist signaling, where the code rate of $\frac{1}{2}$ and the length of the codeword of $N = 672$ are used.
In Fig. \ref{proposed_method_results}, the performance of the proposed detector of the CI-FTN technique is depicted for $\tau = 0.6$, $0.5$, and $0.45$, and is also compared to the no ISI case (i.e., $\tau = 1$) for uncoded and channel coded scenarios. Since CI-FTN signaling with BPSK modulation uses I and Q components, its SE is compared to that in Nyquist signaling with QPSK modulation. At BER of $10^{-5}$, the proposed detector of CI-FTN signaling with BPSK modulation follows the performance of Nyquist signaling with QPSK modulation for $\tau = 0.5$, with the same SE at the same bandwidth and energy despite ISI. At $\tau = 0.45$, the detector achieves $\frac{1.71-1.54}{1.71} = 10\%$ more spectral efficiency and gets close about $0.6$ dB to the no ISI case ($\tau = 1$) at the cost of low complexity, compared to the Nyquist signaling with QPSK modulation. Again at BER of $10^{-5}$, in the channel coded performance that the detector at $\tau = 0.45$ is compared with no ISI, there is about 0.6 dB between them. It shows that channel coding provides a gain similar to Nyquist signaling for CI-FTN signaling. 

\begin{figure}[!t]
\centering
\includegraphics[width=0.55\textwidth]{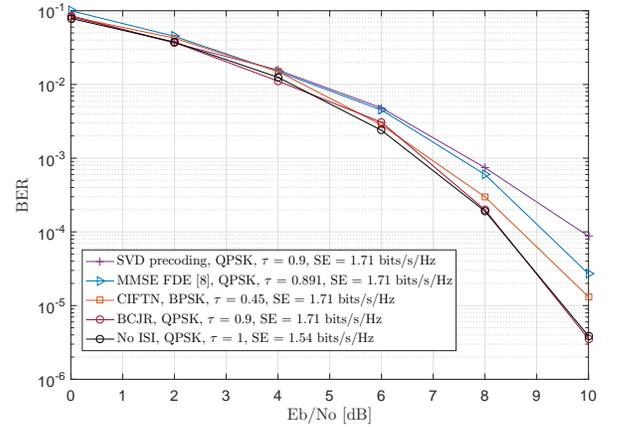}
\caption{Comparison of the proposed detector with different detection algorithms, $\alpha = 0.3.$} 
\label{proposed_method_vs_M_BCJR}
\end{figure}

In Fig. \ref{proposed_method_vs_M_BCJR}, the performance of the low-complexity detector of CI-FTN signaling is compared to the MMSE-FDE in \cite{response_MMSE_FDE}, SVD precoding in \cite{SVD}, and the M-BCJR in \cite{Prlja6241379} with QPSK modulation after setting $\tau$ to the values that result in the same SE of $1.71$ bits/s/Hz for each detector. At the BER of $10^{-4}$, the M-BCJR has the same performance as at $\tau = 1$, but with more complexity. The proposed low-complexity detector gets approximately $0.3$ dB closer to the no ISI case or the M-BCJR compared to the MMSE-FDE with similar complexity to the detector, while the SVD precoding method is about $1.7$ dB away from $\tau = 1$.

\section{Conclusion}\label{sec:conc}

Despite the additional detection complexity, FTN signaling is a promising bandwidth-efficient transmission method that has the potential to be used in next-generation communication systems. In this study, we introduced a component interleaving method called CI-FTN, which leads to constructive ISI between the counter-clockwise rotated BPSK symbols, and presented a corresponding low-complexity FTN detector. The performance and complexity of the proposed detector is compared with standard FTN systems.
The simulation results show that at $\tau=0.5$ the proposed detector of CI-FTN achieves the same QPSK Nyquist signaling performance (no ISI, $\tau=1$). For $\tau=0.45$, the proposed CI-FTN can achieve a gain of $10\%$ in SE compared to QPSK Nyquist at the cost of $0.6$ dB in energy per bit. Hence, CI-FTN provides an interesting trade-off between performance and complexity.


\section*{Acknowledgment}
This work was funded by the Scientific and Technological and Research the Scientifical and Technological Research Council of Turkiye (TUBITAK) Project Grant No. 122E236.


\bibliographystyle{IEEEtran}
\bibliography{IEEEabrv,ciftn_paper.bib}

\begin{thebibliography}{10}
\providecommand{\url}[1]{#1}
\csname url@samestyle\endcsname
\providecommand{\newblock}{\relax}
\providecommand{\bibinfo}[2]{#2}
\providecommand{\BIBentrySTDinterwordspacing}{\spaceskip=0pt\relax}
\providecommand{\BIBentryALTinterwordstretchfactor}{4}
\providecommand{\BIBentryALTinterwordspacing}{\spaceskip=\fontdimen2\font plus
\BIBentryALTinterwordstretchfactor\fontdimen3\font minus
  \fontdimen4\font\relax}
\providecommand{\BIBforeignlanguage}[2]{{%
\expandafter\ifx\csname l@#1\endcsname\relax
\typeout{** WARNING: IEEEtran.bst: No hyphenation pattern has been}%
\typeout{** loaded for the language `#1'. Using the pattern for}%
\typeout{** the default language instead.}%
\else
\language=\csname l@#1\endcsname
\fi
#2}}
\providecommand{\BIBdecl}{\relax}
\BIBdecl

\bibitem{Prlja6241379}
A.~Prlja and J.~B. Anderson, ``Reduced-complexity receivers for strongly
  narrowband intersymbol interference introduced by faster-than-{N}yquist
  signaling,'' \emph{IEEE Transactions on Communications}, vol.~60, no.~9, pp.
  2591--2601, Sep. 2012.

\bibitem{response_MMSE_FDE}
S.~Sugiura, ``Frequency-domain equalization of faster-than-{N}yquist
  signaling,'' \emph{IEEE Wireless Communications Letters}, vol.~2, no.~5, pp.
  555--558, Oct. 2013.

\bibitem{Bedeer7886296}
E.~Bedeer, M.~H. Ahmed, and H.~Yanikomeroglu, ``A very low complexity
  successive symbol-by-symbol sequence estimator for faster-than-{N}yquist
  signaling,'' \emph{IEEE Access}, vol.~5, pp. 7414--7422, 2017.

\bibitem{UNGERBOECK8114271}
S.~Li, B.~Bai, J.~Zhou, P.~Chen, and Z.~Yu, ``Reduced-complexity equalization
  for faster-than-{N}yquist signaling: New methods based on {U}ngerboeck
  observation model,'' \emph{IEEE Transactions on Communications}, vol.~66,
  no.~3, pp. 1190--1204, Nov. 2018.

\bibitem{Liu9458230}
B.~Liu, S.~Li, Y.~Xie, and J.~Yuan, ``A novel sum-product detection algorithm
  for faster-than-{N}yquist signaling: A deep learning approach,'' \emph{IEEE
  Transactions on Communications}, vol.~69, no.~9, pp. 5975--5987, Jun. 2021.

\bibitem{Wang8947975}
H.~Wang, A.~Liu, Z.~Feng, X.~Liang, H.~Liang, and B.~Cai,
  ``Iterative-detection-aided {T}omlinson-{H}arashima precoding for
  faster-than-{N}yquist signaling,'' \emph{IEEE Access}, vol.~8, pp.
  7748--7757, 2020.

\bibitem{Li8954896}
Q.~Li, F.-K. Gong, P.-Y. Song, G.~Li, and S.-H. Zhai, ``Beyond {DVB-S2X}:
  Faster-than-{N}yquist signaling with linear precoding,'' \emph{IEEE
  Transactions on Broadcasting}, vol.~66, no.~3, pp. 620--629, Sep. 2020.

\bibitem{Yuan9352013}
Y.~Li, S.~Xiao, J.~Wang, and W.~Tang, ``Cholesky-decomposition aided linear
  precoding and decoding for {FTN} signaling,'' \emph{IEEE Wireless
  Communications Letters}, vol.~10, no.~6, pp. 1163--1167, Jun. 2021.

\bibitem{Boutros681321}
J.~Boutros and E.~Viterbo, ``Signal space diversity: A power- and
  bandwidth-efficient diversity technique for the {R}ayleigh fading channel,''
  \emph{IEEE Transactions on Information Theory}, vol.~44, no.~4, pp.
  1453--1467, Jul. 1998.

\bibitem{SVD}
H.~Wang, A.~Liu, X.~Liang, S.~Peng, and K.~Wang, ``Linear precoding for
  faster-than-{N}yquist signaling,'' in \emph{Proc. IEEE International
  Conference on Computer and Communications}, 2017, pp. 52--56.

\end{thebibliography}
\end{document}